\documentclass[preprint,twocolumn,10pt,aps]{revtex4-2}

\usepackage{bm}
\usepackage{amssymb}
\usepackage{amsmath}
\newcommand{\TT}{\mathrm{T}}
\newcommand{\HH}{\mathrm{H}}
\usepackage{braket}
\usepackage{graphicx}

\begin{document}

\title{Contour Integral-based Quantum Algorithm for Estimating Matrix Eigenvalue Density}
\author{Yasunori Futamura}
\author{Xiucai Ye}
\author{Tetsuya Sakurai}
\affiliation{Department of Computer Science, University of Tsukuba}


\begin{abstract}
The eigenvalue density of a matrix plays an important role in various types of scientific computing such as electronic-structure calculations.
In this paper, we propose a quantum algorithm for computing the eigenvalue density in a given interval.
Our quantum algorithm is based on a method that approximates the eigenvalue counts by applying the numerical contour integral and the stochastic trace estimator applied to a matrix involving resolvent matrices.
As components of our algorithm, the HHL solver is applied to an augmented linear system of the resolvent matrices, and the quantum Fourier transform (QFT) is adopted to represent the operation of the numerical contour integral.
To reduce the size of the augmented system, we exploit a certain symmetry of the numerical integration.
We also introduce a permutation formed by CNOT gates to make the augmented system solution consistent with the QFT input.
The eigenvalue count in a given interval is derived as the probability of observing a bit pattern in a fraction of the qubits of the output state.
\end{abstract}

\maketitle

\section{Introduction}
The matrix eigenvalue problem is a fundamental linear algebraic problem and has many applications including electronic-structure calculations, vibration analysis, and shell-model calculations.
An efficient method for solving large-scale sparse problems is needed to handle precise computational models of complex phenomena in various scientific fields.

Although the computations of individual eigenvalues are of practical importance, 
the density of the eigenvalues in a certain interval also has a wide variety of applications, such as the density of states in electronic-structure calculations.
The eigenvalue density is also used for the parameter setting in projection methods used to compute the interior eigenvalues~\cite{Sakurai2013}.

The simplest approach to computing the eigenvalue density is first splitting the interval of interest into bins, and then computing the eigenvalue count of each bin.
Therefore, the core of the process is an algorithm for computing the eigenvalue counts within a given interval.


For classical computers, a stochastic estimation method based on a contour-integral was proposed for eigenvalue counts of (non-Hermitian) generalized eigenvalue problems~\cite{Futamura2010}.
A numerical contour integration of the traces of the resolvent matrices that represents a band-pass filter for the spectrum is applied.
For computational efficiency, a stochastic trace estimator is utilized to avoid the direct computation of the elements of the matrix inverses.
This method has been applied to large-scale scientific computing such as shell-model calculations~\cite{Shimizu2016}.

During the last decade, quantum algorithms for linear algebraic computations have emerged owing to the development of the HHL algorithm~\cite{HHL2009} for solving linear systems.
A quantum algorithm for computing a matrix-valued function based on the contour integral has recently been proposed~\cite{Takahira2020}.
The algorithm takes advantage of HHL to solve an augmented linear system derived from a numerical contour integration.

In this paper, we propose a quantum algorithm of the aforementioned contour integral-based method for computing the eigenvalue counts.
The HHL algorithm is utilized to solve an augmented linear system, and the size of the augmented system is reduced by exploiting a certain symmetry of the numerical integration.
The operation of the numerical integration using the trapezoidal rule is straightforwardly represented by the quantum Fourier transform (QFT)~\cite{Nielsen2009} with the help of a permutation applied to the augmented system solution.
The permutation is formed using a group of CNOT gates.
The eigenvalue count in a given interval is estimated based on the probability of observing a certain bit pattern in a fraction of the output qubits of our algorithm.

Throughout this paper, $\|\cdot\|$ denotes the 2-norm of a vector.
Here, $A^\HH$ is the conjugate transpose of matrix $A$.
In addition, $\mathrm{Vec}([\bm{x}_1, \bm{x}_2, \dots, \bm{x}_n])$ returns the vector formed by vertically stacking the column vectors of the input matrix, i.e.,~$[\bm{x}_1^{\HH}, \bm{x}_2^{\HH}, \dots, \bm{x}_n^{\HH}]^{\HH} = \mathrm{Vec}([\bm{x}_1, \bm{x}_2, \dots, \bm{x}_n])$.

The remainder of this paper is organized as follows.
In the next section, we introduce a contour integral-based method for computing the eigenvalue counts and its implementation for classical computers.
Section~\ref{sec:quantum} describes our proposed quantum algorithm.
Finally, we provide some concluding remarks in Section~\ref{sec:conclusion}.

\section{Contour integral-based Stochastic Estimation for Eigenvalue Counts}
\label{sec:classical}

We consider the Hermitian eigenvalue problem
\begin{equation}
	A \bm{q} = \lambda \bm{q},
\end{equation}
where $A \in \mathbb{C}^{n \times n}$ is a complex Hermitian matrix,
$\lambda \in \mathbb{R}$ is an eigenvalue,
and $\bm{q} \in \mathbb{C}^n \backslash \{\bm{0}\}$ is the corresponding eigenvector.
We consider computing the eigenvalue count $m$ in an open interval $\mathcal{I}:=(a, b)$.

In general, $A$ can be diagonalized as 
\begin{equation}
	A = Q \Lambda Q^{\HH} = \sum_{j=1}^n \lambda_j \bm{q}_j \bm{q}_j^{\HH},
\end{equation}
where $\Lambda$ is a diagonal matrix with all eigenvalues, and $Q$ is a unitary matrix whose columns are the eigenvectors.

Let $\Gamma$ be a positively oriented closed curve on a complex plane that intersects the real axis at $a$ and $b$.
Here, we define a projector $P_\Gamma$ as
\begin{align}
	P_\Gamma &:= \frac{1}{2 \pi \mathrm{i}} \oint_\Gamma (z I - A)^{-1} \mathrm{d} z \\
	&= \frac{1}{2 \pi \mathrm{i}} \sum_{j=1}^n \oint_\Gamma \frac{1}{z - \lambda_j} \mathrm{d} z \bm{q}_j \bm{q}_j^{\HH}\\
	&= \sum_{j=1}^{n} g(\lambda_j) \bm{q}_j \bm{q}_j^{\HH}
\end{align}
where $\mathrm{i} := \sqrt{-1}$, $I$ denotes the $n$-dimensional identity matrix, and
\begin{equation}
	g(\lambda) := \frac{1}{2 \pi \mathrm{i}} \oint_\Gamma \frac{1}{(z - \lambda)} \mathrm{d}z.
\end{equation}
Through Cauchy's integral expression, we have
\begin{equation}
	g(\lambda) =
	\begin{cases}
 		1 & \mathrm{if} \quad \lambda \in \mathcal{I},\\
 		0 & \mathrm{otherwise}.
	 \end{cases}
\end{equation}
Therefore,
\begin{align}
	P_\Gamma &= \sum_{j \mid \lambda_j \in \mathcal{I}} \bm{q}_j \bm{q}_j^{\HH} \\
	&= Q I_\Gamma Q^{\HH},
\end{align}
where $I_\Gamma$ is a diagonal matrix whose $j$-th diagonal element is $g(\lambda_j)$.
Because the trace of the matrix is invariant under any similarity transformation,
$\mathrm{Tr}(P_\Gamma) = \mathrm{Tr}(I_\Gamma)$ gives the eigenvalue count in $\mathcal{I}$.



To approximate the contour integration, we use the numerical quadrature
\begin{equation}
	\widetilde{P}_\Gamma := \sum_{k=1}^N w_k (z_k I - A)^{-1},
\end{equation}
where $z_j$ and $w_j$ are quadrature points and quadrature weights.
In practice, a quadrature rule that satisfies
\begin{equation}
	z_k = \overline{z_{N-k-1}}, w_k = \overline{w_{N-k-1}} \quad (k = 0,1,\dots,N/2-1).
	\label{eq:conj}
\end{equation}
is preferred to take advantage of the symmetry of $A$.
Thus, we assume that~\eqref{eq:conj} holds.

The eigenvalue count $m$ can be approximated by
\begin{align}
	\mu &:= \mathrm{Tr}(\widetilde{P}_\Gamma) \\
	&= \mathrm{Tr} \left(\sum_{k=0}^{N-1} w_k (z_k I - \Lambda)^{-1} \right) \\
	&= \sum_{j=1}^n \sum_{k=0}^{N-1} \frac{w_k}{z_k- \lambda_j}\\
	&= \sum_{j=1}^n f_N(\lambda_j),
\end{align}
where
\begin{equation}
	f_N(\lambda) := \sum_{k=0}^{N-1} \frac{w_k}{z_k - \lambda}.
\end{equation}
Here, $f_N$ is regarded as an approximation of $g$ and is regarded as a band-pass filter for the spectrum.
The behavior of the filter function is analyzed in~\cite{Maeda2015}.

To avoid the direct computation of the trace involving the matrix inverses,
we consider utilizing a stochastic estimation of the matrix trace described below.

Let $M \in \mathbb{R}^{n \times n}$ be a general matrix, and $\bm{v} \in \mathbb{R}^n$ be a random vector whose mean is the zero vector and covariance matrix is the identity.
It is known that the mean of the quadratic form involving $M$ and $\bm{v}$ gives $\mathrm{Tr}(M)$~\cite{Provost1992}, namely,
\begin{equation}
	\mathbb{E}(\bm{v}^\TT M \bm{v}) = \mathrm{Tr}(M),
\end{equation}
where $\mathbb{E}(\cdot)$ indicates the mean.
This is easily extended for the general complex matrix.

We define
\begin{equation}
	\bm{x}_k := (z_k I - A)^{-1} \bm{v}
	\label{eq:xk}
\end{equation}
and
\begin{align}
	\bm{s} &:= \widetilde{P}_\Gamma \bm{v} \\
	&= \sum_{k=0}^{N-1} w_k \bm{x}_k.
\end{align}
Therefore, $\mu$ is represented by
\begin{equation}
	\mu = \mathbb{E}(\bm{v}^{\HH} P_\Gamma \bm{v}) = \mathbb{E}(\bm{v}^{\HH} \bm{s}).
\end{equation}
The eigenvalue count is estimated by computing the arithmetic mean of $\bm{v}^{\HH} \bm{s}$ with samples of the random vector $\bm{v}$.

Because of the resolvent matrix in~\eqref{eq:xk}, the most computationally dominant part is to solve linear systems
\begin{equation}
	 (z_k I - A) \bm{x}_k = \bm{v} \quad (k = 0,1,\dots,N-1).
	\label{eq:linsys}
\end{equation}
For classical computers, Krylov subspace methods are utilized to solve the systems~\cite{Futamura2010}.

\section{Quantum Algorithm}
\label{sec:quantum}

The estimator $\mu$ can be used in general non-Hermitian cases~\cite{Futamura2010}.
By exploiting the symmetry of $A$, we derive another formulation of estimating the eigenvalue count.
The formulation is more compatible with our quantum algorithm described later.
Hereafter, we assume that $n$ and $N$ are powers of 2 such that $n = 2^{b_n}$ and $N = 2^{b_N}$.

By computing
\begin{equation}	
\nu := \mathrm{Tr}({\widetilde{P}_\Gamma}^2),
\end{equation}
we can also approximate $m$ because
\begin{align}
	\mathrm{Tr}({\widetilde{P}_\Gamma}^2) &= \sum_{j=1}^n f_N(\lambda_j)^2.
\end{align}
When $f_N$ is a good approximation of $g$, $f_N(\lambda)$ is small where $\lambda$ is distant from $\mathcal{I}$.
In such a case, ${f_N}^2$ is also a good approximation unless $f_N$ oscillates excessively in $\mathcal{I}$.

Because $A$ is Hermitian and \eqref{eq:conj} holds, it can be easily shown that $\widetilde{P}_\Gamma$ is also Hermitian.
Using the trace estimator, $\nu$ is estimated by
\begin{align}
	\mathrm{Tr}({\widetilde{P}_\Gamma}^2) &= \mathbb{E}(\bm{v}^{\HH} {\widetilde{P}_\Gamma}^2 \bm{v}) \\
	&= \mathbb{E} ( \bm{v}^{\HH} {\widetilde{P}_\Gamma}^{\HH} \widetilde{P}_\Gamma \bm{v} ) \\
	&= \mathbb{E} (\|\bm{s}\|^2).
\end{align}
Therefore, by sampling $\|\bm{s}\|^2$ and computing the arithmetic mean, we can estimate the eigenvalue count.

To derive a quantum algorithm, we specifically assume that $\Gamma$ is a circle with center $\Gamma \in \mathbb{R}$ and radius $\rho > 0$, and consider using the trapezoidal rule for the numerical quadrature.
Through this setting, $z_k$ and $w_k$ are given as
\begin{equation}
	z_k = \gamma + \rho w_k
\end{equation}
and
\begin{equation}
	w_k = \frac{\rho}{N} e^{\frac{2 \pi \mathrm{i} \left( k + \frac{1}{2} \right) }{N}} \quad (j=0,\dots,N-1).
	\label{eq:wk}
\end{equation}
This quadrature rule satisfies \eqref{eq:conj}.
Note that the term $\frac{1}{2}$ in~\eqref{eq:wk} is introduced to ensure that $(z_j I - A) \quad (j=0,\dots,N-1)$ are non-singular (all eigenvalues of $A$ are real).
In addition, $f_N$ with the trapezoidal rule on a circle does not oscillate in $\mathcal{I}$.
Thus, it is preferable to be used to compute $\mu$.

Here, we define quantum states $\ket{v} := \bm{v}/ \|\bm{v}\|$ and $\ket{s} := \bm{s}/ \|\bm{s}\|$.
We propose a quantum algorithm to estimate $\|\bm{s}\|$, whose mean gives an approximation of the eigenvalue count.

Herein, we define
\begin{align}
	\bm{t}_j &:= \frac{1}{\sqrt{N}} \sum_{k=0}^{N-1} e^{\frac{2 \pi \mathrm{i} jk}{N}} \bm{x}_k.
\end{align}
We then have $\bm{s} = \frac{\rho e^{\frac{\pi \mathrm{i}}{N}}}{\sqrt{N}} \bm{t}_1$ and
\begin{equation}
	\|\bm{s}\| = \frac{\rho}{\sqrt{N}} \|\bm{t}_1\|.
	\label{eq:sandt1}
\end{equation}
We show an algorithm to derive the quantum state $\ket{t_1} := \bm{t}_1/\|\bm{t}_1\| = \ket{s}$ using HHL and QFT.

We consider solving linear systems~\eqref{eq:linsys} by forming an augmented linear system and using HHL as in a quantum algorithm for a contour integral-based matrix function operation~\cite{Takahira2020}.
Instead of the symmetrization used for HHL~\cite{HHL2009},
we apply a different form by taking advantage of a certain symmetry of the problem.

Let $A_k := z_k I - A$ and $A^\prime := A_0 \oplus A_1 \dots \oplus A_{N/2-1}$,
$\bm{v}^\prime := \mathrm{Vec}([\bm{v}, \bm{v}, \dots, \bm{v}]) \in \mathbb{R}^{nN}$.
\begin{equation}
C: = 
\begin{pmatrix}
O & A^{\prime} \\
{A^{\prime}}^{\HH} & O \\
\end{pmatrix},
\end{equation}
where $O$ is the $n$-dimensional square zero matrix.
Note that $C$ is Hermitian.
The size of this augmented linear system is half that in the symmetrization form used for HHL.
Because \eqref{eq:conj} and $A = A^{\HH}$ hold,
\begin{equation}
	(z_{N-k-1} I - A) = (z_k I - A)^{\HH} \quad (0,1,\dots,N/2-1).
\end{equation}
We therefore have
\begin{equation}
	C \bm{y} = \bm{v}^{\prime},
	\label{eq:augment}
\end{equation}
where
$\bm{y} := \mathrm{Vec} ([\bm{x}_0, \bm{x}_1, \dots, \bm{x}_{N/2-1}, \bm{x}_{N-1}, \bm{x}_{N-2}, \dots, \bm{x}_{N/2}] )$.
By solving the augmented linear system~\eqref{eq:augment} using HHL, we obtain the quantum state $\ket{y} := \bm{y}/ \|\bm{y}\|$, which can be represented as
\begin{equation}
\begin{split}		
	 \ket{y} &= \frac{1}{\|\bm{y}\|} \sum_{k=0}^{N/2-1} \left( \|\bm{x}_k\| \ket{k} \ket{x_k} \right. \\
	 &+ \left. \|\bm{x}_{N - 1 - k}\| \ket{N/2 + k} \ket{x_{N - 1 - k}} \right),
\end{split}
\label{eq:y1}
\end{equation}
where $\ket{x_k} := \bm{x}_k / \|\bm{x}_k\|$.
To obtain a quantum state proportional to $\bm{y}^\prime :=\mathrm{Vec} ([\bm{x}_0, \bm{x}_1, \dots, \bm{x}_{N-1}] )$,
we consider a unitary operator that transforms \eqref{eq:y1} into
\begin{equation}
	 \ket{y^\prime} := \frac{1}{\|\bm{y}\|} \sum_{k=0}^{N-1} \|\bm{x}_k\| \ket{k} \ket{x_k}.
	 \label{eq:yprime}
\end{equation}
Such a unitary operator can be formed with a permutation $\Pi$ of $\{0,1,\dots,N-1\}$
\begin{equation}
	\Pi(k) =
	\begin{cases}
 		k & \mathrm{if} \quad k < N/2\\
 		3N/2 - k - 1& \mathrm{otherwise}.
	 \end{cases}
\end{equation}
This permutation can be implemented by 2-qubit CNOT gates controlled by the most significant bit.

\begin{figure}[t]
   \centering
   \includegraphics[width=8.5cm]{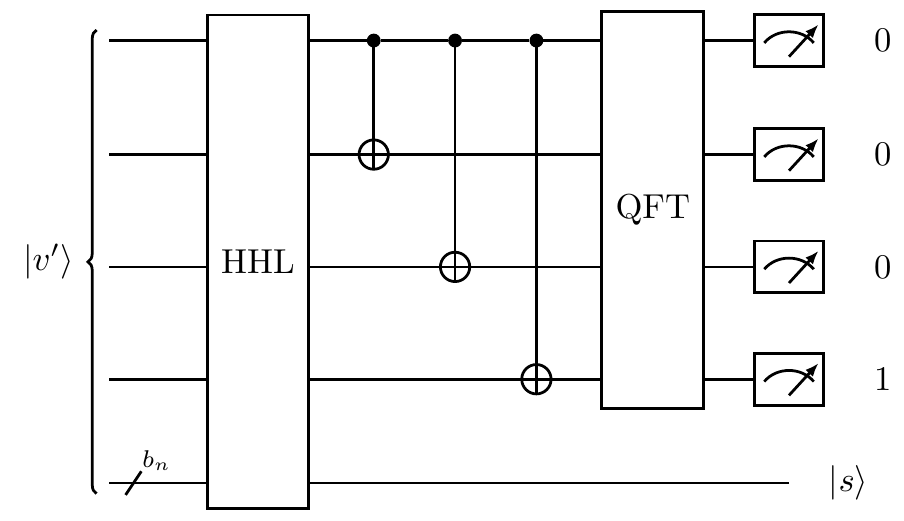}
   \caption{The quantum circuit of our algorithm where the number of quadrature points $N= 2^4$.}
   \label{fig:circuit}
\end{figure}

By applying $U_\mathrm{QFT} \otimes I$ to $\ket{y^\prime}$, we have
\begin{align}		
	\ket{\psi} &:= (U_\mathrm{QFT} \otimes I) \ket{y^\prime} \\
	&=\frac{1}{\sqrt{N} \|\bm{y}\|} \sum_{j=0}^{N-1} \ket{j} \left( \sum_{k=0}^{N-1} e^{2 \pi \mathrm{i} jk} \| \bm{x}_k \| \ket{x_k} \right) \\
	&= \frac{1}{\|\bm{y}\|} \sum_{j=0}^{N-1} \| \bm{t}_j \| \ket{j} \ket{t_j}.
\end{align}
Consequently, when we observe $00\dots01$ at the significant $b_N$ qubits, we obtain a quantum state $\ket{s} = \ket{t_1}$.
The probability $p$ of observing $00\dots01$ is 
\begin{equation}
	p = \frac{\|\bm{t}_1\|^2}{\|\bm{y}\|^2}.
\end{equation}
Because of~\eqref{eq:sandt1}, $p$ can be expressed as
\begin{equation}
	p = \frac{N \|\bm{s}\|^2}{\rho^2 \|\bm{y}\|^2}.
\end{equation}
Therefore, through $p$ and $\|\bm{y}\|$, we can estimate $\|\bm{s}\|^2$ whose mean gives $\mu$, which is an approximation of the eigenvalue count.
In addition, by applying a swap test for $\ket{s}$ and $\ket{v}$, we can estimate $\bm{v}^{\HH}\bm{s}$ to compute $\nu$, which is an alternative to $\mu$.
Note that $\|\bm{y}\|^2$ can be estimated by the probability of observing 0 on the auxiliary bit of the HHL.
The resulting quantum circuit of our algorithm is shown in FIG.~\ref{fig:circuit}.

%

\section{Conclusion}
\label{sec:conclusion}
In this study, we proposed a quantum algorithm for estimating the eigenvalue density.
Our algorithm is based on the eigenvalue count computation using the numerical contour integration and a stochastic matrix trace estimator.
As the main component, the HHL algorithm is utilized to solve an augmented linear system involving the resolvent matrices derived from the numerical contour integration.
In our formulation, the size of the augmented linear system is reduced by exploiting the symmetry of $A$ and the quadrature rule.
The QFT is utilized to represent the operation of the trapezoidal rule as a numerical quadrature.
To make the augmented system solution consistent with the QFT input, we formulate a permutation based on CNOT gates.
The resulting eigenvalue count is obtained through the probability of observing a certain bit pattern in a fraction of the qubits of the output state.

\bibliography{ref}

\end{document}